\documentclass[a4paper,11pt]{article}
\usepackage{tabularx,graphicx,moreverb}
\usepackage{amsmath,amssymb,amsthm}
\usepackage{bm}
\usepackage{ascmac}
\usepackage{makeidx}
\usepackage[english]{babel}
\usepackage{wrapfig}
\usepackage{color}
\usepackage{algorithm}
\usepackage{setspace}
\usepackage{algorithmic}
\usepackage[round]{natbib}
\usepackage[margin=1in]{geometry}
\begin{document}

\newcommand{\ra}{\rightarrow\,}
\newcommand{\sst}{\shortstack\,}
\theoremstyle{break}
\newtheorem{lem}{Lemma}
\newtheorem{them}{Theorem}
\newtheorem{asmp}{Assumption}

\title{An approximation method for discrete Markov decision models with a large state space}
\author{Masaaki Imaizumi
\\University of Tokyo}

\date{\today}

\thispagestyle{empty}

\maketitle

\begin{abstract}
We propose a new approximation approach to solve a \textit{discrete Markov decision model} (DMD) with a large state space.
The DMD is an structural model which can analyze data obtained from agents making dynamic decisions, however, to solve DMDs with a large number of discrete states is always difficult (and at times, impossible) because of a huge computational cost.
The number of the states in DMDs increases exponentially as we introduce state variable, and this phenomenon is called ``The Curse of Dimensionality.''
To overcome this problem, we propose the new approach, named a \textit{statistical least square temporal difference method} (SLSTD), that can solve DMDs containing the large state space with a low computational cost.
The SLSTD can easily solve a Bellman equation of DMDs with a high dimensional variable, by employing two approximation techniques.
Experimentally, the SLSTD performs faster and more accurate than other existing methods, and in some cases, reduces the computation time by over 99 percent.
We also show that an estimator for a parameter of interest obtained by the SLSTD has the consistency and the asymptotically normality.


\end{abstract}


\section{Introduction}

A \textit{discrete Markov decision model} (DMD), also known as a dynamic discrete choice model, is extensively used for analyzing a behavior of agents.
The main advantage of the approach with the DMD is that it admits us to implement the counterfactual analysis of agents, since the DMD can handle the dynamic decision making of the agents.
The agents in the DMD observe their own state in each period, then decide their action with considering future reward and transition between states.
The action by the agents is often formalized as a discrete choice, and we can analyze the characteristics of the agents by investigating the realized choice.
\citet{Rust1987} suggested the DMD, and has found many applications in present-day works in econometrics, marketing science, transportation science, and the dynamic games theory.

Since the estimator for the parameter of the DMD rarely has an analytical form, implementing the DMD approach requires numerical calculation.
However, to solve the DMDs requires conducting the highly complex nonlinear computation, hence the computational cost restricts the flexibility of the DMD.
Thus, many empirical researches are suffered from the computational hurdle and forced to reduce volume of their DMD.
Accordingly, many methodological researches have suggested computational methods to remove the hurdle.
\citet{HM1993}, \citet{AM2002}, \citet{Su2012}, and \citet{DFS2012} proposed methods to solve the DMD with a single agent.
\citet{AM2007}, \citet{BBL2007}, \citet{PS2008}, and \citet{ELS2015} proposed efficient methods to solve the DMD including multiple agents and game structures.

``The curse of dimensionality,'' that refers to the exponential rise in the number of grid points in the state space of the DMD, is one of the such computational difficulties.
When we increase the number of discretized state variables in the DMD, the grid points in the state space increases exponentially and it leads to the extreme rise in the computation cost.
Furthermore, such large state spaces require a large amount of computational memory to store a great number of numerical values, and in some instances, ordinary computers are not even up for the task.

Several methods have been suggested to handle to the curse of dimensionality in the DMD.
There are some general methods that we can apply into various DMDs.
\citet{KW1997} and \citet{Imai2004} proposed a method to approximate the value function by basis functional approximation method.
A series estimation based on \citet{judd1996} is also a method to handle the curse and it can be applicable to a wide range of DMDs.
There are other methods for specific DMDs which are specialized to analyze specific topics.
\citet{HN2006} and \citet{Gowrisankaran2012} solve specific DMDs for analyzing consumer choices.
Another method is the Monte Carlo method by \citet{Rust1997} that can solve any model with up to a certain number of dimensions.

Despite the rich researches against the curse of dimensionality, few methods achieve both the generality and the sufficient cost reduction.
Some methods are valid for specific problem, like \citet{HN2006} and \citet{Gowrisankaran2012}, thus it is not applicable when we try to analyze the other topics.
On the other hand, the general methods such as  \citet{KW1997} do not have enough theoretical analysis which guarantees the result of the analysis, and also their performance is not sufficient in some cases.
(We will discuss its detail in Section 4).

Purpose of this paper is to suggest a new computational method which is applicable for wide range of DMDs and has sufficient computational and statistical performance.
In this paper, we suggest a \textit{statistical least square temporal difference method} (SLSTD) that can avoid the curse of the dimensionality of the DMD.
The SLSTD focus on the Bellman equation of the DMD.
Solving the Bellman equation is an origin of the computational burden when the state space is large, and reducing the cost of handling the Bellman equation is a critical problem.
The SLSTD simplify the Bellman equation by applying basis function approximation to a high dimensional variable of the Bellman equation.
Furthermore, the SLSTD employs the stochastic root-finding technique to solve the simplified Bellman equation.
By the two techniques, we can substantially reduce the computational burden from the curse of dimensionality.
The combination of the two methods is based on the idea of the temporal difference (TD) method by \citet{Sutton1988} and the least square temporal difference (LSTD) method by \citet{BB1996}: we extend both methods for applying them to the DMD models.


Our numerical experiments reveal advantages of the SLSTD over approximation method by \citet{KW1997} and the series estimation.
First, the SLSTD can provide more accurate results of the parameter estimation.
Second, the computation time by the SLSTD is nearly independent of the size of the state space, and as such, the computation time remains small even when the state space is large.
Third, the SLSTD also has an advantage from an aspect of the computational memory.
We also provide the asymptotic properties of the parameter estimation obtained by the SLSTD.
Given some conditions on the smoothness of models and the number of basis functions, we can even obtain consistency and asymptotic normality.

The rest of paper is organized as follows.
Section 2 describes the fundamental structure of the DMD and the existing approximation methods to handle the curse of dimensionality. Section 3 introduces the SLSTD. Section 4 examines performance of the SLSTD by numerical experiments. Section 5 shows the theoretical aspects of the SLSTD. Section 6 concludes. The proofs are collected in the Appendix.

\section{Model and existing methods}

\subsection{Model}

The DMD is a statistical model to analyze a sequence of discrete choices, and the purpose of the analysis is to estimate the parameter of the agents from their choices. 
In the DMD, the observed choices depend on the state in which the agent stays, and the action of the agents determine the transition between states. 
We derive the likelihood of the actions, and estimate the parameters of interest by maximizing the likelihood.

We consider the DMD with discrete time and the discrete state variable.
DMDs are formulated as $(\mathcal{S},\mathcal{A},\Theta,u,\mathcal{P})$. $\mathcal{S}$ is a state space with $p$ dimensions, and each dimension $j$ has $q_j$ states. 
Then, the state space is represented as $\mathcal{S} = \{ s_{11},...,s_{1q_1} \} \times ... \times \{ s_{p1},...,s_{pq_p} \}$ and $|\mathcal{S}| = \prod _{j=1}^p q_j$. 
$\mathcal{A}$ is an action space, and $\Theta \subset \mathbb{R}^d$ is a parameter space. $u : \mathcal{S} \times \mathcal{A} \times \Theta \times \mathcal{E} \to \mathbb{R}$ is a reward function, and $\mathcal{E}$ is the space of stochastic factors.
$\mathcal{P} : \mathcal{S} \times \mathcal{A} \times \mathcal{S} \to [0,1]$ is the transition probability between states. $s'$ represents the state in the next period.

There exist $n$ agents, and an agent $i = 1,\ldots, n$ observes own state $s_{i,t} \in \mathcal{S}$ in each period $t$.
Then the agent $i$ decide own action $a_{i,t} \in \mathcal{A}$ from the action space.
Also, the agent $i$ privately observes $\epsilon_{i,t}$ which is an independent stochastic factors and it is unobservable for the researchers.
The state of the agents $i$ evolves after the action $a_{i,t}$ has been made.
Also it is assumed that the transition between states has a first order Markov property.

Denote by $\theta \in \Theta$ the parameter vector which explains the characteristic of whole agents, and $\theta$ is the parameter of interest for the researchers.
We set that the agent $i$ obtains reward $u(s_{i,t'},a_{i,t'},\epsilon_{i,t'};\theta)$ with the stochastic factor $\epsilon_{i,t}$ and the parameter vector $\theta$.
At time $t$ and with given the state $s_{i,t}$, the agent $i$ maximizes the discounted sum of the reward, named \textit{the value function} as follows,
\begin{align}
	v(s_{i,t};\theta) &:= \max_{\{a_{i,t} \in \mathcal{A}\}_{t}}E\left[\left. \sum _{t' = t} \beta^{t' - t} u(s_{i,t'},a_{i,t'},\epsilon_{i,t'};\theta) \right| s_{i,t} \right], \label{value_func}
\end{align}
where $\beta \in [0,1)$ is the discount factor and $s_0$ is the initial state.
Here, $v(s;\theta) : \mathcal{S} \times \Theta \rightarrow \mathbb{R}$ is a function, and also define $V(s_{i,t};\theta) := E[v(s_{i,t};\theta)]$ and $V(\theta) = \{V(s; \theta)\}_{s \in \mathcal{S}}$

To analyze the decision making of the agent $i$, we consider the choice probability of the action in each period.
Let $P(a|s_{i,t};\theta,V(\theta))$ be the probability of choosing $a_{i,t}$ in state $s_{i,t}$ with given parameter $\theta$:
\begin{align}
	P(a|s_{i,t};\theta,V(\theta)) := E\left[ 1(a =\arg \max_{\tilde{a}}\{ u(s_{i,t},\tilde{a}, \epsilon_{i,t};\theta) + \beta E_{s' \sim P}[v(s';\theta)|s_{i,t},\tilde{a}] \}) \right], \label{cond_prob}
\end{align}
where $1(\cdot)$ is an indicator function and the expectation $E_P[\cdot]$ is taken over the state transitions $\{p(s_{i,t+1} | s_{i,t}, s_{i,t})\}_t$.

Since $\mathcal{A}$ has finite elements, we can combine \eqref{value_func} and \eqref{cond_prob}, then we obtain the following equation, named \textit{the Bellman equation}, as follows, 
\begin{align}
	V(s_{i,t};\theta) =& \sum _{a \in \mathcal{A}} P(a | s_{i,t};\theta,V(\theta)) \left[ U(s_{i,t},a,\epsilon_{i,t};\theta) + \beta \sum_{s' \in \mathcal{S}}V(s';\theta)p(s'|s_{i,t},a) \right], \label{disc_bellman}
\end{align}
where $ U(s_{i,t},a,\epsilon_{i,t};\theta)$ is an expectation of $u(s_{i,t},a,\epsilon_{i,t};\theta)$ with given $a$.
By solving the Bellman equation, we can obtain the value of $V(\theta)$.

Suppose that we observe a sequence of state transitions and actions $\{(s_{i,t},a_{i,t})\}_{i,t=1}^{n,T}$. 
The likelihood of the observed sequence of derived by the conditional choice probability \eqref{cond_prob}, when we obtain the following log likelihood function with observation $\{(s_i,a_i)\}_{i=1}^n$:
\begin{align}
	\mathcal{L}(\theta, V(\theta)) := \frac{1}{nT}\sum _{i=1}^{n}\sum _{t=1}^{T} \log p(s_{i,t+1}|s_{i,t},a_{i,t}) + \log P(a_{i,t}|s_{i,t};\theta,V(\theta)). \label{criteria}
\end{align}
By maximum likelihood estimation, we obtain the estimator $\hat{\theta} := \arg \max \mathcal{L}(\theta)$ while $V(\theta)$ satisfies the Bellman equation \eqref{disc_bellman}.

\paragraph{Remark}

Practically, calculating the value of $U(s_{i,t},a,\epsilon_{i,t};\theta)$ and $P(a | s_{i,t};\theta,V(\theta))$ requires a tedious numerical integration.
To avoid the computation, several assumptions on the functional and distributional form of $u(s,a,\epsilon ; \theta)$ are often introduced.

When we are allowed to assume that $u(s,a,\epsilon;\theta) = \overline{u}(s,a;\theta) + \epsilon_a$ where $\epsilon_a$ is a stochastic term that is i.i.d. with respect to action $a$ and time, and $\epsilon _a$ follows the type-I extreme value distribution, we obtain the following simple forms:
\begin{align*}
	P(a|s;\theta,V(\theta)) = \frac{\exp(\overline{u}(s,a;\theta) + \beta E_P[V(s';\theta)|s,a])}{\sum _{\tilde{a}}\exp(\overline{u}(s,\tilde{a};\theta) + \beta E_P[V(s';\theta)|s,\tilde{a}])}.
\end{align*}
In this case, the Bellman equation (\ref{disc_bellman}) can be rewritten as
\begin{align*}
	V(s;\theta) =& \sum _{a} P(a | s;\theta,V(\theta))\left[ \overline{u}(s,a;\theta) + E[\epsilon _{a} |s, a;\theta,V(\theta)] + \beta \sum _{s'} f(s'|s,a) V(s';\theta) \right], 
\end{align*}
where  $E[\epsilon_{a} |s, a;\theta,V(\theta)]$ represents a conditional expectation of $\epsilon _{a}$, i.e.,
\begin{align*}
	E[\epsilon _{a} |s, a;\theta,V(\theta)] = \log \left[ \frac{\sum _{\tilde{a}}\exp(\overline{u}(s,\tilde{a};\theta)) + \beta E_P[V(s';\theta)|s,\tilde{a}])}{\exp(\overline{u}(s,a;\theta) + \beta E_P[V(s';\theta)|s,a])}\right] + \gamma, 
\end{align*}
where $\gamma$ is the Euler's constant. 
This form enables us to calculate $U(s_{i,t},a,\epsilon_{i,t};\theta)$ and $P(a | s_{i,t};\theta,V(\theta))$ analytically.

\citet{Rust1987} introduced the assumptions, and have drastically reduced the cost of numerical integration to calculate transition probabilities. It is discussed in detail in \citet{AM2002}.

\subsection{The Curse of Dimensionality and Existing Methods}

To solve the Bellman equation \eqref{disc_bellman} is necessary to evaluate the log likelihood function in \eqref{criteria}, however, it is quite difficult when the state space $\mathcal{S}$ is large.
Since the state variables are discretized, the Bellman equation \eqref{disc_bellman} is regarded as an equation of $|S|$-dimensional vector.
Namely, let $\textbf{V} =\left( (V(s))_{s \in \mathcal{S}}\right)^T \in \mathcal{R}^{|\mathcal{S}|}$, and rewrite the Bellman equation as $\textbf{V} = \mathfrak{T}(\textbf{V})$, where $\mathfrak{T}$ is the right hand side of \eqref{disc_bellman}.
However, as the number of the state variables $p$ increases, the number of states is $|\mathcal{S}| = \prod _{j =1}^{p} q_{j}$ exponentially increases against $p$, and solving the equation $\textbf{V} = \mathfrak{T}(\textbf{V})$ requires huge computational time and cost.
For instance, in the DMD for the career decision, if we allow each agents to possess $6$-types of the human capitals for maximum $40$ years for each of the types, then we obtain $|\mathcal{S}| = 40^6 = 4,096,000,000$ in the DMD and we have to solve an equation with $4,096,000,000$-dimensional vector.
It requires a huge computational time, and also note that ordinal laptops cannot contain such the high-dimensional vector in their computational memory.
Similar examples are also introduced in several literature, for example, \citet{HN2006} for the consumer choice and \citet{ELS2015} for the discrete choice game.

To avoid the curse, there exist some methods to solve the DMD under the curse of dimensionality; these are of two types: general methods and problem-specific methods.

\citet{KW1994} suggested a method that can be applied to general types of DMDs.
This method picks some states randomly in each time, and estimates the coefficients of an interpolation function of the picked states as
\begin{align*}
	&E[ \max _a \{u(s,a) + \beta E[V(s')]\}] \\
	&\approx \psi( \max _a E[u(s,a) + \beta E[V(s')]], \mbox{mean}_a E[u(s,a) + \beta E[V(s')]] ),
\end{align*}
where $\psi(\cdot,\cdot)$ is the interpolation function.
Though it is handy, this method has some faults. First, the computation time increases rapidly.
The method is mainly suited for simplifying Bellman equation evaluation, and thus is not good at reducing the state-space computation cost. Second, the theoretical framework of the method has not been sufficiently elaborated on.
Since performance is guaranteed only by numerical experiments, its theoretical properties, such as consistency and size of biases, are unknown.

We also consider another general method using the sequential series estimation method, which can solve many DMDs.
This method, too, picks states from the state space in each time, and approximates $V(s;\theta)$ as
\begin{align*}
	V(s_t;\theta) \approx \sum _k r_{\theta,tk} \phi_k(s_t),
\end{align*}
where $r_{tk}$ is a weight and $\phi_k(s_t)$ is a basis function. 
Since the method approximates the value function in each period, the method requires multiple approximations. \citet{judd1996} provides the idea of the series estimation, and this method applied the idea to DMDs.
This method is useful and its convergence is theoretically guaranteed, but it does have one limitation, which we discuss later. \citet{Rust1997} too suggested a general method with Monte Carlo that can calculate a value function with no effect of an increase in the number of dimensions.
Their method, though independent of the object of analysis, requires strong restrictions on the transition and state space of the model.

Some problem-specific methods, such as those by \citet{HN2006} and \citet{Gowrisankaran2012}, are for consumer choices. 
These methods display high performance in market analysis, but depend on the specific characteristics of the market and are not applicable to other DMDs.

Thus, while problem-specific methods are fast, they cannot solve other general problems, and while some general methods enjoy wide applicability, they are not computationally feasible.
Accordingly, there is a need for a general method that achieves computational feasibility.

Other popular method to estimate the DMD include the conditional choice probability (CCP) method by \citet{HM1993}, the nested pseudo likelihood (NLP) method by \citet{AM2002}, and the mathematical programming with equilibrium constraint (MPEC) method by \citet{Su2012}. The SLSTD works under the curse of dimensionality, whereas these methods are design to solve DMDs with relatively small state spaces.
As an example, consider the carrier decision model by \citet{KW1997}, which has over 1 million states ($|\mathcal{S}| > 1,000,000$). It is impossible for implement the MPEC and CCP methods to solve such a model, because these methods need to provide a $|\mathcal{S}| \times |\mathcal{S}|$ numerical matrix, which is computationally not feasible.
\section{Proposed Method}

We introduce the SLSTD which solves the Bellman equation \eqref{disc_bellman} approximately with low computational cost.
The SLSTD employs mainly two techniques, (i) the functional approximation method, and (ii) the stochastic approximation method.
After solving the Bellman equation by the SLSTD, we provide a formation about (iii) the parameter estimation.
The main idea of the SLSTD is based on the TD method by \citet{Sutton1988} and the LSTD method by \citet{BB1996} and \citet{Nedic2003}. 

Preliminarily, we provide some notation.
$\|f\|_{\mathcal{S}}^2 := \sum_{s \in \mathcal{S}} f(s)^2$. 
Let $\{\phi_j(s) \}_{j=1}^{\infty}$ is an orthonormal system in $L^2(\overline{\mathcal{S}})$ with $\phi_j : \overline{\mathcal{S}} \rightarrow \mathbb{R}$, where $\overline{\mathcal{S}} \subset \mathbb{R}^p$ is a convex hull of $\mathcal{S}$.
With given $k > 1$, consider a vector-valued function $\phi(s) := (\phi _1(s),...,\phi _k(s))^T$.

For brevity, we define a Bellman operator $T[\cdot]$ such as
\begin{align}
	T[V(\theta)](s,a) := U(s,a,\epsilon) + \beta \sum_{s' \in \mathcal{S}}V(s';\theta)p(s'|s,a), \notag
\end{align}
then the Bellman equation (\ref{disc_bellman}) can be rewritten as
\begin{align}
	V(s;\theta) = \sum_{a \in \mathcal{A}} P(a|s;\theta,V(\theta))T[V(\theta)](s,a). \label{bellman_simple}
\end{align}
Also we let $V^*(\theta)$ and $P^*(\theta,V^*(\theta))$ be a solution of the Bellman equation.
Note that the part $(V(\theta)^*, P(\theta,\theta^*))$ is probed to be unique by \citet{RTW2002} for each $\theta$.

\subsection{Method Outline}

In this section, we provide an outline of the SLSTD.
Purpose of the SLSTD is to solve the Bellman equation \eqref{bellman_simple} (the simplified version of \eqref{disc_bellman}).
The SLSTD employs the two approximation techniques, (i) the basis functional approximation, and (ii) the stochastic approximation method.

\paragraph{(i) Basis Functional Approximation of the Value Function :}

With given $\theta$, we approximate the value function $V(s;\theta)$ as
\begin{align}
	V(s;\theta) \approx \phi^T(s) w_{\theta} = \sum_{j=1}^k \phi_j(s) w_{\theta,j}, \label{func_approx}
\end{align}
where $w \in \mathbb{R}^k$ is a vector of approximation weights $w_{\theta,j}$. 
Since the approximation is regarded as a projection of $V(s)$ onto the linear space spanned by $\{\phi_j(s)\}$, there exists a unique optimal weight $w^*_{\theta} := \arg\min_{w_{\theta}} \| \phi(\cdot)^T w_{\theta} - V^*(\cdot;\theta)\|_{\mathcal{S}}^2$ by the projection theorem.
Note that $\min_{w_{\theta}} \| \sum_{j=1}^k \phi_j(\cdot) w_{\theta,j} - f(\cdot)\|_2^2$ converges to zero as $k \rightarrow \infty$ when $f$ is sufficiently smooth (See \citet{Tsybakov2009}).  

By the functional approximation \eqref{func_approx}, we can represent $V(s;\theta)$ ($|\mathcal{S}|$-dimensional vector) by $w_{\theta}$ ($k$-dimensional vector) by using the given orthonormal system.
Since we set $k$ is much less than $|\mathcal{S}|$, we can avoid the high dimensionality of the Bellman equation.

However, it is not enough to solve the curse of dimensionality.
There are some problems remain : (a) the problem of obtaining $w_{\theta}^*$ which approximately solves the Bellman equation remains, (b) the computational cost reduction is not enough, and (c) the accumulation of the approximation problem appears.
Especially, the problem (a) is critical.
The solution of the Bellman equation should satisfy the equation \eqref{func_approx} for all $s \in \mathcal{S}$.
Thus, it is necessary for evaluating the equation \eqref{func_approx} for all $s \in \mathcal{S}$ to solve the Bellman equation by ordinal method, such as the Newton's method.
However, as we already discussed, $|\mathcal{S}|$ is too large in some cases, hence it require a high computation cost. 
In the rest of the section, we introduce additional method to solve the problem (a).
The problems (b) and (c) will be discussed in Section 4.

\paragraph{(ii) Stochastic Approximation for Obtaining $\hat{w}_{\theta}$ :}

To estimate the optimal weight $w^*_{\theta}$, we evaluate the goodness of the approximation with given $w_{\theta}$.
In the view of $\sum_{a \in \mathcal{A}}P(a|s;\theta,V(\theta)) = 1$, $V^{*}(\theta)$ satisfies $\sum_a P(a|s;\theta,V(\theta))\left[ V(s) - T[V(\cdot)](s,a)\right] = 0$ for all $s \in \mathcal{S}$ with given $\theta$.
Then, we define the similar moment condition for $w_{\theta}$ as follows.
We consider the minimization problem
\begin{align}
	\min _{w_{\theta}} \sum _{s \sim D} \left[ \sum _{a \in \mathcal{A}}  P\left(a|s;\theta,\phi^T(s) w_{\theta}\right) \{ \phi^T(s) w_{\theta} - T[\phi^T(\cdot) w_{\theta}](s,a) \}\right]^2,\label{opt_prob2}
\end{align}
where $D$ is a set of $s \in \mathcal{S}$ generated from the empirical distribution.
For the minimization problem, we implement Lemma 6 in \citet{TR1997}, we obtain the weak form of a first order condition \eqref{opt_prob2}.
\begin{align}
	 \sum _{s \sim D}\phi(s) \left[ \sum _{a \in \mathcal{A}} P\left(a|s;\theta,\phi^T(s) w_{\theta}\right) \left( \phi^T(s) w_{\theta} - T[\phi^T(\cdot) w_{\theta}](s,a) \right)\right] = 0. \label{foc}
\end{align}

To solve \eqref{foc}, we implement \textit{the stochastic approximation method} in \citet{Benveniste2012}.
The stochastic approximation method is an algorithm to find a root of an equation which is given by a form of an expectation.
As a sequence of random variables generated from a probability distribution is observed one-by-one, the stochastic approximation update the solution of the equation, and the sequence of the solution converges to the root.
Here, we cite Theorem of the stochastic approximation method from \citet{Benveniste2012}.

\newtheorem*{thm_stochapp*}{Theorem: Stochastic approximation (\citet{Benveniste2012}, Theorem 17)}

\begin{thm_stochapp*}
Let $X$ be a random variable with transition probability $\Pi(x',x)$ and denote $F(w) = E_{\Pi}[f(w,X)]$ where $E_{\Pi}[\cdot]$ is a expectation under stationary distribution. Suppose that there exists a unique $\xi^*$ satisfying $F(w^*) = 0$. Further, consider a decreasing sequence $\{\eta_i\}_i$, where $\sum _i^{\infty} \eta _i = \infty$ and $\sum _i^{\infty} \eta _i^2 < \infty$. Suppose
\begin{itemize}
\item $f(w,x)$ has an envelope function with a polynomial of $x$ and a linear function of $w$
\item $(w - w^*)E_{\Pi}[f(w,X)] < 0$ holds for all $w \neq w ^*$.
\end{itemize}
hold. If a sequence of $\{w_i\}$ is generated in the following iteration equation 
\begin{align*}
	w _{i+1} = w _i + \eta_i f(w_i,X_i),
\end{align*}
then the sequence $\{w_i\}$ converges to $w^*$ with probability $1$ as $i \ra \infty$.\footnote{We set $U(w) = (w - w_0)^2$ and $\rho_n(w,X_i) = 0$ in Theorem 17 of \citet{Benveniste2012}.}
\end{thm_stochapp*}

To apply the theorem, we set $X = s$, $f(w_{\theta},X_i) = \phi(s) \sum _{a \in \mathcal{A}} P\left(a|s;\theta,\phi^T(s) w_{\theta}\right) \{T[\phi^T(\cdot) w_{\theta}](s) - \phi(s)^Tw_{\theta}\}$ and $\Pi(X_i,X_{i+1}) = f(s'|s,a) P(a|s;w_{\theta})$. 
We check the assumptions in Appendix. 
This algorithm by the stochastic approximation method enables us to solve the problem \eqref{foc} without summing up or integrating $\phi(s) \sum _a P\left(a|s;\theta,\phi^T(s) w_{\theta}\right) \{T[\phi^T(\cdot) w_{\theta}](s) - \phi(s)^Tw_{\theta}\}$ with respect to the all state in each step.

By the stochastic approximation method, we define the sequence $\{w_{\theta}^{(\ell)}\}_{\ell = 1}^{\infty}$ from the following equation:
\begin{align}
	w_{\theta}^{(\ell + 1)} = w_{\theta}^{(\ell)} + \eta _{\ell} \phi(s_{\ell}) \sum _{a \in \mathcal{A}} P\left(a|s;\theta,\phi^T(s) w_{\theta}^{(\ell)}\right) \left(T[\phi^T(\cdot) w_{\theta}^{(\ell)}](s_{\ell},a) - \phi(s_{\ell})w_{\theta}^{(\ell)}\right), \label{iter_stoch}
\end{align}
with step size $\eta_{\ell}$, which satisfies $\sum _{\ell = 1}^{\infty} \eta _{\ell} = \infty$ and $\sum _{\ell}^{\infty} \eta _{\ell}^2 < \infty$. 
The initial point of $w_{\theta}^{(0)}$ is arbitrary.
The basic approach underlying the SLSTD is modifying the approximation parameter $w_{\ell}$ iteratively as per the temporal difference; this is why we refer to the method as the TD method.

Based on the stochastic approximation method, we define the estimator of $w_{\theta}^*$ as a limit of the sequence \eqref{iter_stoch} as follows:
\begin{align*}
	\hat{w}_{\theta} := \lim_{\ell \rightarrow \infty} w_{\theta}^{(\ell)}.
\end{align*}

Now, by the SLSTD approach, we obtain the estimator of the value function $V(s;\theta)$ as
\begin{align*}
	\hat{V}(s;\theta) := \phi^T(s)\hat{w}_{\theta},
\end{align*}
with given $\theta$.
Note that the estimation of $V(s;\theta)$ is implemented for each fixed $\theta$, and the estimator $\hat{V}(s;\theta)$ can differ for each $\theta$.

\paragraph{(iii) Parameter Estimation : }

Finally, we define the estimator of the parameter of interest by the SLSTD as
\begin{align*}
	\hat{\theta} := \arg \max_{\theta} \mathcal{L}(\theta, \hat{V}(\theta)),
\end{align*}
where $\mathcal{L}$ is defined in \eqref{criteria}.
Note that when we have to evaluate the value of $\mathcal{L}(\theta, \hat{V}(\theta))$ with different $\theta$ during the optimization with respect to $\theta$, we have to rerun the SLSTD for each $\theta$.
It looks costly at first glance, however, it is not a problem in practice.
Details are shown in Section 4.

\subsection{Implementation and Discussion}

To proceed with iteration (\ref{iter_stoch}), we need to prepare the sequence of the states $\{s_{\ell}\}_{\ell}$.
Since we use the set $D$ of the state from the empirical distribution in \eqref{opt_prob2}, we use the state transitions from the observed data for the iteration.
In other words, The SLSTD approach intensively minimizes the error of the Bellman equation \eqref{bellman_simple} on states whose agents pass frequently.

Consider $n$ agents, with the $i$th agent having a state transition of length $T$. 
Now we have a set of observed state transition $\{ s_{i,t} \}_{i,t = 1}^{n,T}$. 
First, we implement the iteration (\ref{iter_stoch}) on the state transition of one agent. 
When the first agent reaches a terminal state, we carry on the $w$ and continue the iteration (\ref{iter_stoch}) with the state transition of the next agent. 
We repeat the operation for all $n$ agents. 
Thus, total number of observation is $N = nT$. 
This method fits structure of data in econometric field which has many agents, contrast to the ordinal TD method implements the iteration with one long state transition.

This operation has another interpretation. 
We unite the data of all agents as one agent's repetitive action to generate the decision sequence. 
If the model has a terminal state and the repetitive agent reaches it, he does not obtain any reward and goes back to the initial state with probability one. 
Then, the state transition become irreducible for all state, thus we can recognize the state transition as having stationary distribution.

Practically, we do not have to generate the sequence $\{w_{\theta}^{(\ell)}\}_{\ell}$ until $\ell = N$.
We can stop the iteration, when we judge the convergence of the sequence, namely, $\|w_{\theta}^{(\ell)} - w_{\theta}^{(\ell + 1)}\|$ is no less than the sufficient small predetermined tolerance level $\tau > 0$.

We now provide a pseudo code of the SLSTD. Algorithm \ref{pcode} shows how to solve the Bellman equation with given parameter $\theta$. A sequence of $\eta _{\ell}$ is predetermined, and along with the initial approximation weight $w_{\theta}^{(0)}$. 

\begin{algorithm}
	\caption{SLSTD : Derive $\hat{w}_{\theta}$ with given $\theta$}
	\label{pcode}
	\begin{algorithmic}
		\STATE \textbf{Given} $\theta$, $\{ s_{i,t} \}_{i,t = 1}^{n,T}$, $\{\phi_j(s)\}_{j=1}^k$,$\tau > 0$,$\{ \eta _{\ell} \}_{\ell=1}^{\infty}$ s.t. $\sum _{\ell=1}^{\infty} \eta _{\ell}= \infty$ and $\sum _{\ell=1}^{\infty} \eta _{\ell}^2 < \infty$
		\STATE \textbf{Initialize} $w \leftarrow w_{\theta}^{(0)}$ arbitrary
		\STATE \textbf{Set} $t = 1$
		\STATE $\ell \leftarrow 1$
		\LOOP
			\FOR{$i = 1, \ldots , n$}
				\FOR{$t = 1, \ldots , T_k$}
					\STATE $s \leftarrow s_{i,t}$
					\IF{$s$ is the terminal state}
						\STATE $w_{\theta}^{(\ell + 1)} \leftarrow w_{\theta}^{(\ell)} + \eta _{\ell} \phi(s)\{ - \phi^T(s) w_{\theta}^{(\ell)}\}$
					\ELSE
						\STATE $w_{\theta}^{(\ell + 1)} \leftarrow w_{\theta}^{(\ell)} + \eta _{\ell} \phi(s) \sum_{a \in \mathcal{A}} P\left(a|s;\theta,\phi^T(s) w_{\theta}^{(\ell)}\right) \left(T[\phi^T(\cdot) w_{\theta}^{(\ell)}](s,a)- \phi^T(s) w_{\theta}^{(\ell)}\right)$
					\ENDIF
					\STATE $\ell \leftarrow \ell + 1$
				\ENDFOR
			\ENDFOR
			\IF{$\|w_{\theta}^{(\ell + 1)} - w_{\theta}^{(\ell)}\| \leq \tau$}
				\STATE break loop
			\ENDIF
		\ENDLOOP
	\end{algorithmic}
\end{algorithm}

We now discuss the initial point of $w_{\theta}^{(0)}$ and the tuning parameter $\{\eta_{\ell}\}_i$. When $w_{\theta}^{(0)}$ is far from the limit, the convergence of $w_{\theta}^{(\ell)}$ becomes unstable. 
Thus, the initial point needs tuning when the solution is not stable. 
The select of the step size is a more important problem. We use $\eta_{\ell} = \frac{c_1}{\ell + c_2}$ satisfies $\sum _{\ell=1}^{\infty} \eta_{\ell}= \infty$ and $\sum _{\ell = 1}^{\infty} \eta _{\ell}^2 < \infty$ with positive $c_1$ and $c_2$. When $c_1$ is too small or $c_2$ is too large, the step size becomes small and solution is strongly affected by the initial point. Thus, selecting a proper step size, too, is necessary to obtain a stable solution.

We now discuss the choice of basis functions. The power series and B-splines are common as basis functions. However, the B-spline function provides better estimates. Since the SLSTD locally evaluates the value function approximation on the observed state transition, the B-spline functions provide more precise approximation.
\section{Numerical Experiment}\label{nume_sec}

\subsection{Parameter Estimation}

To compare the estimation accuracy of the methods, we conducted a numerical experiment using some empirical DMDs. 
We consider the method of \citet{KW1994} (henceforth, KW), the sequential series estimation, and the SLSTD. 
We used a DMD for analyzing the career decision which is a simplified version of \citet{KW1997}. 
The DMD is a finite time model, with an adjustable number of state variables and actions of the agents, and the time horizon.

We present the DMD used in our experiment and it is often used in the labour economics analysis, such as \citet{KW1997}.
The model has $p$ state variables and its terminal time is $T$. 
An elements of the state space has a form $s = (s_1,s_2,s_3,s_4) \in \mathcal{S}$: $s_1$ is age, $s_2$ is a education year, $s_3$ is a carrier of work and $s_4$ contains the choice in the previous period. 
The state space is constructed as $\mathcal{S} = \{s_{11},...,s_{1T}\} \times \{s_{21},...,s_{2T}\} \times \{s_{31},...,s_{3T}\} \times \{s_{41},...,s_{43}\}$. We set a action space as $\mathcal{A} = \{1,2,3\}$. We set a reward function as
\begin{align*}
	u(s,a;\theta) =
	\begin{cases}
		\theta _1 s_2 ,&\mbox{if}~a = 1, \\
		\theta _2 s_2 + \theta _3 s_3,&\mbox{if}~a=2,\\
		\theta _4,&\mbox{if}~a=3.
	\end{cases}
\end{align*}
The choice $a = 1$ is a decision of schooling and the choice increases $s_1$ and $s_2$ by one. 
The choice $a=2$ is a decision of working, and it increases $s_1$ and $s_3$ by one.
The choice $a = 3$ is staying home, and it increases only $s_1$. 

We generate data with $n = 1,000$ agents, and estimate the parameters by the generated data. 
$q = |\mathcal{S}|$ represents the number of elements in the state space.
When we $T = 30$, then we have $q = 81,000$ elements. 
We set the stochastic term $\epsilon$ to follow a type-I extreme value distribution. 
Since the state space of this model is not so large, we can estimate $\theta$ without the approximation method, such as the SLSTD, the KW method, and the series representation, for comparison.
To use the SLSTD, we use the B-spline functions as the basis functions. 
We also set same basis functions for the sequential series method, and we set $5$ grid points in each dimension for approximation. 
For the KW method, we provide $100$ grid points in each period.

First, we compare the parameter estimation. 
For estimation, we use the simulated numerical data generated from the model with the true parameter. 
We generate data with $1,000$ agents, and replicate the experiment $200$ times. 
Table \ref{estimation1} and Table \ref{estimation2} show the result with different parameter sets. 
The tables also contain the computational time for solving the Bellman equation with one parameter set, and the squared error of the Bellman equation which is a difference between the left hand side and the right hand side of \eqref{bellman_simple}. 
Here, $\Delta$ denotes the error of the Bellman equation.
The values are the means of the estimators of the replications; the figures in parentheses are standard deviations. 

The SLSTD provides a less estimation bias than other methods for every cases. 
In some cases, the sequential series estimator has better performance; this is because the model is linear and very simple, and thus, is likely to perform better in series estimations. 
However, when model gets larger or complex, the sequential series method does not work well. 
In contrast, the standard deviation of the SLSTD are larger than others. 
The KW method performs unstable for most cases. 
As the state space bigger, the bias becomes larger. 

From a point of the Bellman error, the SLSTD performs best, and the series method performs worst. From a point of computational cost, the KW methods requires more computational time. The SLSTD and the sequential series method provides less computational cost in Table \ref{estimation1} and Table \ref{estimation2}. We will investigate this point further in the following section.

\subsection{Computational Cost}

Next, we see the results of experiments about computational burden. 
We use the same model as in the estimation part, and change the value of $p$ and $T$. 
The following results show the time-to-solve the Bellman equation for given true parameters. 
For estimation, we repeat the same process for 200 times.

Table \ref{time_all} shows the results. 
The unit of time is seconds. 
We can see that the sequential and the KW methods cannot avoid the exponentially increasing computational burden. 
The burden is particularly severe for the KW method. 
In contrast, the computation cost by the SLSTD does not increase exponentially. 

We provide some explanations for these results. 
First, the SLSTD evaluates the Bellman equation on less number of states. 
The SLSTD refers only the state that is observed as data, and avoid checking other states due to the stochastic approximation method. 
In contrast, other methods need to refer more number of states for evaluating the Bellman equation. 
Secondly, the SLSTD uses less computational memory, thus it realizes less computational cost. 
The sequential series and the KW methods require a large memory to store the numerical values of $\{V(s;\theta)\}_{s \in \mathcal{S}}$. 
When the memory for use is quite large, accessing the memory becomes much costly. 
On the other hand, the SLSTD only stores information about the approximation weight $w_{\theta}$ and the basis functions. 
When the values of some states are required, the weight and basis functions are sufficient to recall $\{V(s;\theta)\}_{s \in \mathcal{S}}$. 
This contributes to the computational advantage of the SLSTD.

Note also that the sequential series method fails to obtain the value when $p = 5$ and $T = 30$. 
This is because of the increased error from the accumulation of sequential approximation. 
Because it accepts backward induction, we have to approximate $\{V(s;\theta)\}_{s \in \mathcal{S}}$ in each period. 
As $\{V(s;\theta)\}_{s \in \mathcal{S}}$ is approximated several times, the approximation error accumulates, and sometimes the accumulated error diverges. 
Figure \ref{errseq} shows the error accumulation. 
We try to approximate the DMD with labor economics in Section \ref{nume_sec} using the sequential series method. 
The horizontal axis is the time period, and the vertical axis shows the value of the approximated $V(s,\theta)$. 
We can observe the error accumulation and that its size rises exponentially by the multiple approximation. 

When $p = 5$ and $T = 40$, the KW method and the sequential series method cannot yield any results. 
This is because the computation time is so long that we cannot obtain the estimation results. 
Since the state space is quite large in this case, the computational memory in laptops cannot handle the numerical value of $\{V(s;\theta)\}_{s \in \mathcal{S}}$ in the usual way : in other words, these two methods are not appropriate. 
In contrast, the SLSTD only stores the values of the weight $w_{\theta} \in \mathbb{R}^k$, and it does not need a large memory to store the value of $\{V(s;\theta)\}_{s \in \mathcal{S}}$.

\subsection{Discussion}

In this section, we discuss the advantages of the SLSTD. 
First, the SLSTD does not suffer from computer memory limitations. 
As mentioned earlier, the sequential series method and the KW methods require a large memory to store $|\mathcal{S}|$ numerical values, and it is at times, impossible to store all values in the state space. 
For instance, when $|\mathcal{S}| = 10^{10}$, the requirement is 80 GB of memory.
In contrast, the SLSTD only stores value of $w$ which can recover value of $V(s;\theta)$ for all $s$. 

\begin{figure}[htbp]
 \begin{center}
  \includegraphics[width=150mm]{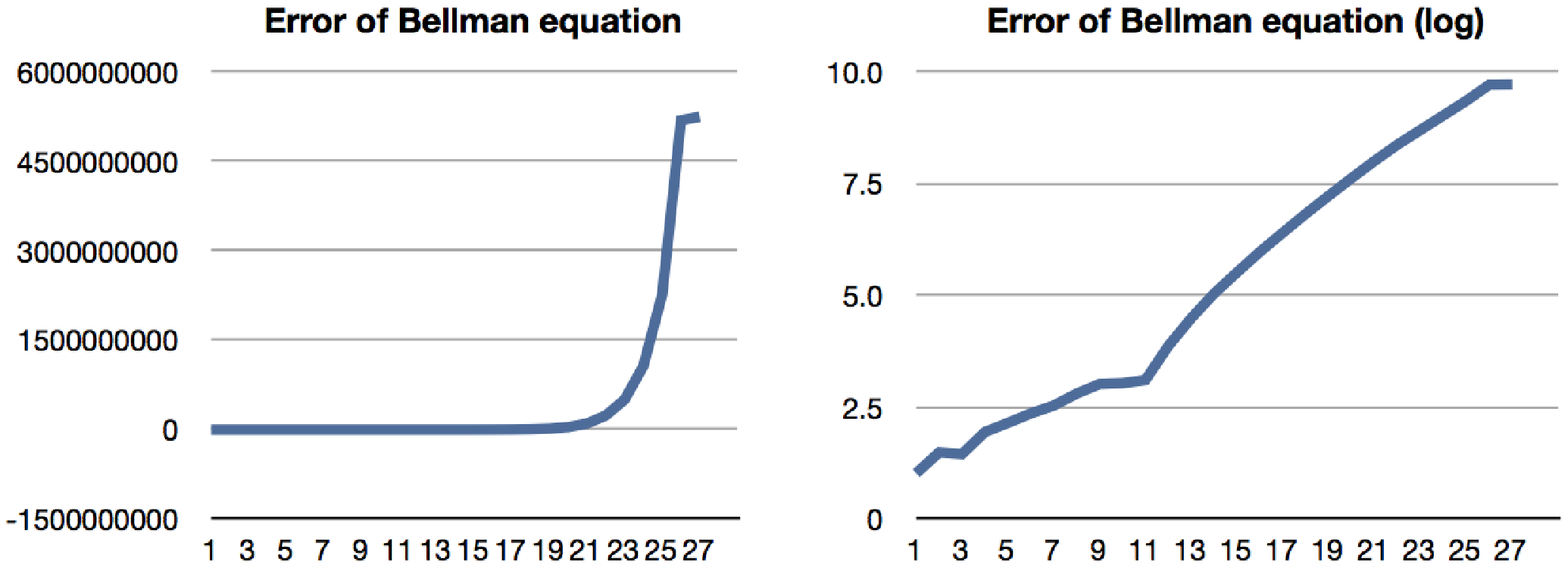}
 \end{center}
 \caption{Error accumulation for the sequential series method}
 \label{errseq}
\end{figure}

The second advantage is in terms non-smoothness of DMDs. 
Figure \ref{errseq} presents a accumulation of approximation error by the series method, when the DMD has non-smooth term. 
As the series method requires multiple approximation, the horizontal axis of figure \ref{errseq} a number of the approximation and the vertical axis is a size of the error of the Bellman equation. 
It is easy to check that the multiple approximation causes the error approximation. 
The sequential series method fails when $V(s;\theta)$ does not have enough smoothness. 
In contrast, the SLSTD can avoid the problem because it approximates the $V(s;\theta)$ at once. 
In addition, the SLSTD delivers a theoretical analysis about smoothness and estimation. 
In following section, we provide a theoretical analysis of the SLSTD.

The third advantage lies when applying to a non-logit type model. 
Throughout many empirical researches, it is often required that the stochastic term $\epsilon$ has a type-I extreme distribution and the choice probability is represented in the multinomial logit form. 
If we use a non-logit type model, the derivation of choice probability requires costly numerical integration. 
However, as shown earlier, the SLSTD reduces the number of times the choice probability needs to be derived. 
Thus, the SLSTD has a relative advantage when applying to a non-logit type model.



\section{Theory}

In this section, we show the consistency of $\hat{V}(s;\theta)$ provided by the SLSTD, and the asymptotic properties of the estimator $\hat{\theta}$. 
In this section, we define the norm $||\cdot||$ as the Euclid norm, and consider $||F(\cdot)||_{\mathcal{S}} = \sup _{s \in \mathcal{S}}|F(s)|$. 
$\nabla _x$ represents a partial differentiation with respect to $x$.

\subsection{Property of $\hat{V}(s;\theta)$}

As for evaluating $\hat{V}(s;\theta)$, our theoretical results of the stochastic approximation part mainly depends on \citet{TR1997} and \citet{Tagorti2014}. The error in the approximation by the basis functions comes from nonparametric series estimation, in the line of \citet{Newey1997} and \citet{Andrews1991}.


We also consider a following asymptotic setting¥ :  $q = |\mathcal{S}|$ increases as $N$ increases. 
We write the settings as $q \geq C N$ for some $C > 0$. 
In the field of empirical researches, there is a correlation between the size of state space and the number of observation. 
For example, in the DMD we used in Section \ref{nume_sec}, $N$ and $q$ are correlated through terminal time $T$. 
This setting decently explains properties of the state spaces in actual empirical researches.

To apply the theories, we assume the following conditions. 
$E_0[\cdot]$ denotes the expectation with a stationary distribution. 
To evaluate the shape of $V(s;\theta)$, we define a new function on continuous space $\dot{V}^* : \overline{\mathcal{S}}\times \Theta \ra \mathcal{R}$ which satisfies $\sup_{s \in \mathcal{S}}|V^*(s;\theta) - \dot{V}^*(s;\theta)| = 0$ for all $\theta$. 
\footnote{An existence of $\dot{V}^*(s;\theta)$ is obvious. If $\dot{V}^*(s;\theta)$ is not unique, we accept $\dot{V}^*(s;\theta)$ with maximum $m$.}

\begin{asmp}
Assume that
\begin{enumerate}
\item $u(s,a,\epsilon;\theta)$ is bounded.
\item $\dot{V}^*(s;\theta)$ is $m$ times differentiable with respect to $s \in \overline{\mathcal{S}}$.
\end{enumerate}
\label{asm_conv}
\end{asmp}

The following lemma provides a consistency of the stochastic approximation method.

\begin{lem} \label{convlem}
If assumption \ref{asm_conv} holds, then for all $\theta \in \Theta$, we obtain with large probability,
\begin{align*}
	||\phi^T(s)\hat{w}_{\theta} - V^*(s;\theta)||_{\mathcal{S}} = O_P\left(\frac{1}{\sqrt{N}}\log N + \frac{k}{\sqrt{N}} +k^{\frac{1}{2}-\frac{m}{p}}\right).
\end{align*}
\end{lem} 
The proof is in Appendix \ref{proof_conv}. Now, we obtain the convergence rate of $\hat{V}(s;\theta)$, as it becomes an important factor of the following asymptotic estimation analysis. When $N,k \ra \infty$, $k^2/N \ra 0$ and $m>2p$, we obtain $||\phi^T(s)\hat{w}_{\theta} - V^*(s;\theta)||_{\mathcal{S}} = o_P(1)$.

\subsection{Asymptotic properties of the estimation}

In this section, we provide the asymptotic normality of the estimator for $\theta$. 
Throughout this analysis, we recognized $V(s;\theta)$ as a nuisance parameter, and treated $\theta$ as a parameter of interest. 
The asymptotic result is provided by \citet{Kosorok2008} which analyzes the semiparametric M-estimator. 

First, we formalize the estimation problem. 
Since that we observe $N = n T$ state transitions, we use $j = 1,\ldots,N$ as an index of the observation, and $i(j)$ and $t(j)$ is a corresponding index.
According to the empirical estimation equation \eqref{criteria}, we rewrite the log likelihood function as
\begin{align*}
	l(Z_j;\theta,V(\theta)) := \log  p(s_{i(j),t(j)+1}|s_{i(j),t(j)},a_{i(j),t(j)}) + \log P(a_{i(j),t(j)}|s_{i(j),t(j)};\theta,V(\theta)),
\end{align*}
where $Z_j := (s_{i(j),t(j)},a_{i(j),t(j)},s_{i(j),t(j)+1}) \in \mathcal{Z} := \mathcal{S} \times \mathcal{A} \times \mathcal{S}$. 
Then, $\mathcal{L}(\theta,V(\theta)) := \frac{1}{N}\sum _{j=1}^{N} l(Z_j;\theta,V(\theta))$ and $\mathcal{L}_0(\theta,V(\theta)) = E[l(Z;\theta,V(\theta))]$. 
We define the true parameter as $\theta_0 = \arg \max_{\theta} \mathcal{L}_0(\theta,V^*(\theta))$.

To show the asymptotic result, we consider the following assumptions, in the line of IL.

\begin{asmp}
\label{asm_asym}
Assume that
\begin{enumerate}
	\item $\theta _0$ is an interior point in the compact $\Theta$, and is the unique maximizer of $\mathcal{L}_0(\theta,V^*(\theta))$. \label{asmp2_1}
	\item $\left. \nabla_{\theta}^2 \mathcal{L}_0(\theta,V(\theta))\right|_{\theta = \theta_0} + \left.\nabla_{V}^2 \mathcal{L}_0(\theta_0)\right|_{V(\theta) = V^*(\theta_0)}$ exists and it is non-singular. \label{asmp2_3}
	\item For all $s \in \mathcal{S}$, $V^*(s;\theta)$ is Lipschitz continuous with respect to $\theta$. \label{asmp2_2}
	\item With some radius $\delta_N = O(N^{-1/4})$, the class $\{l(Z;\theta,V(\theta)) : \mathcal{Z} \rightarrow \mathbb{R} \left| \theta \in \Theta , \|V(\theta) - V^*(\theta_0)\| \leq \delta_N \right. \}$ is P-Donsker.\label{asmp2_4}
	\item $|\nabla \mathcal{L}(\theta, V(\theta)) - \nabla \mathcal{L}(\theta_0, V^*(\theta_0))| = O(\|\theta - \theta_0\|) + O(\|V(\theta) - V^*(\theta_0)\|^2)$ hold as $N \rightarrow \infty$,  for all $\theta \in \Theta , \|V(\theta) - V^*(\theta_0)\| \leq \delta_N$ with some radius $\delta_N = O(N^{-1/4})$. \label{asmp2_5}
\end{enumerate}
\end{asmp}



Assumption 2-\ref{asmp2_1} is an identification condition for the true parameter $\theta_0$.
Assumption 2-\ref{asmp2_3} is for the regularity for the estimation problem of DMDs, and they are generally assumed in the asymptotic statistics. (For example, see \citet{vdV2000}.)
One can criticize that there exist empirical researches without the identifiability of the true parameter, thus we have to pay attention to the such cases.
Assumption 2-\ref{asmp2_2} requires 
Assumption 2-\ref{asmp2_4} is somewhat abstract, however, it can cover the wide range of the functions, such as smooth, monotone, Lipschitz continuous functions, introduced in \citet{vdV2000}.
When we let the reward function $u(s,a;\theta)$ and $p(s'|s,a)$ satisfy the such properties, Assumption 2-\ref{asmp2_4} holds.
Assumption 2-\ref{asmp2_5} requires a kind of smoothness.
Though it is strong assumption a little, it is general requirement in the literature of the semiparametric statistics.


The following theorem provides asymptotic normality.

\begin{them}
\label{asymthm}
If assumptions \ref{asm_conv} and \ref{asm_asym} hold, then
\begin{align*}
	\| \hat{\theta} - \theta_0 \| \stackrel{p}{\longrightarrow } 0,
\end{align*}
as $N \ra \infty$.

Furthermore, if $\frac{k}{\sqrt{N}} + k^{\frac{1}{2}-\frac{m}{p}} = O(N^{-1/4})$ holds, then
\begin{align*}
	\sqrt{N}(\hat{\theta} -\theta_0) \stackrel{d}{\longrightarrow } \mathcal{N}(0,H_0^{-1} E[\Gamma_0 \Gamma_0^T] H_0^{-1}),
\end{align*}
where
\begin{align*}
	\Gamma _0=& E \left[ \left(\nabla _{\theta} \left. l(Z;\theta,V^*(\theta)) \right|_{\theta = \theta_0}+ \nabla _{V} \left. l(Z;\theta,V^*(\theta)) \right|_{\theta = \theta_0}\right)^{\otimes 2} \right]\\
	H_0 =& \nabla_{\theta}^2 \left. \mathcal{L}_0(\theta,V^*(\theta))\right|_{\theta = \theta_0} + \nabla_{\theta}\nabla_{V} \left. \mathcal{L}_0(\theta,V^*(\theta))\right|_{V(\theta) = V^*(\theta_0),\theta = \theta_0} ,
\end{align*}
as $N \ra \infty$.
\end{them}
Here, $Y^{\otimes}$ denotes $Y^T X$ with a vector $Y$. The proof is provided in Appendix \ref{proof_asym}.\footnote{The variance components $\Gamma_0$ and $H_0$ have analytical forms when $\epsilon$ has some parametric distribution.}
This theorem shows that the consistency of $\theta$ is guaranteed in most cases. Moreover, when the smoothness of $V(s;\theta)$ is enough relative to the number of dimensions $p$, we obtain asymptotic normality of the estimator. For instance, $k,N \ra \infty, k = O(N^{1/4})$ and $\frac{m}{p} \geq \frac{3}{2}$ ensures that the condition holds. In the proof, we treat the estimation problem (\ref{criteria}) as a semiparametric M-estimation by regarding $V(s;\theta)$ as a nuisance parameter. \citet{IL2010} enables us to derive the asymptotic variance of the estimator.

\section{Conclusion}
We suggested a new approximation technique, the SLSTD, to solve discrete Markov decision models with a large state space. Because the curse of dimensionality makes the computation cost enormous, it prevents development of research using DMD models. We numerically show that the SLSTD can approximate and solve the Bellman equation with a low computation cost. Further, the asymptotic theory guarantees that the SLSTD has good properties.

\newpage
\appendix
\section{Figures and Tables}

\begin{table}[htbp]
\begin{center}
  \begin{tabular}{c||ccc|cccc|cc}
Method & $p$ & $T$ & $|\mathcal{S}|$ & $\theta_1$ & $\theta_2$ & $\theta_3$ & $\theta_4$ & Time (sec) & $\Delta^2$ \\ \hline \hline
True param &  &  &  & 1.0 & 2.0 & 1.0 & 9.0 & &\\ \hline
SLSTD & 4 & 10 & 3000 & 0.97 & 2.49 & 1.16 & 5.89 & 0.44 & 2.2E+01\\
 &  &  &  & (0.17) & (0.23) & (0.17) & (0.31) & (0.003) & (2.9E+01)\\
Sequential & 4 & 10 & 3000 & 1.06 & 2.03 & 0.99 & 8.91 & 0.09 & 3.0E+04\\
 &  &  &  & (0.04) & (0.11) & (0.02) & (0.18) & (0.001) & (4.1E+04)\\
KW & 4 & 10 & 3000 & -3.49 & 2.25 & 0.21 & 4.63 & 0.64 & 3.5E+02\\
 &  &  &  & (9.66) & (1.51) & (0.29) & (1.74) & (0.007) & (7.8E+02)\\ \hline
SLSTD & 4 & 15 & 10125 & 0.97 & 2.49	& 1.16	& 5.89 & 0.69 & 1.6E+02\\
 &  &  &  & (0.17)	& (0.23)	& (0.17) & (0.31) & (0.004) & (2.5E+02)\\
Sequential & 4 & 15 & 10125 & -0.19 & -0.03 & -0.05 & -5.71 & 0.16 & 4.8E+05\\
 &  &  &  & (0.12) & (0.11) & (0.13) & (7.19) & (0.017) & (7.1E+05)\\
KW & 4 & 15 & 10125 & 4.46 & 3.86 & -0.04 & 3.46 & 2.85 & 1.4E+03\\
 &  &  &  & (2.30) & (2.09) & (0.30) & (2.65) & (0.016) & (2.8E+03)\\ \hline
SLSTD & 4 & 20 & 24000 & 0.58 & 0.88	& 0.64	& 8.13 & 0.95 & 1.3E+03\\
 &  &  &  & (0.70) & (0.81)	& (0.62)	& (6.44) & (0.007)  & (2.3E+03)\\
Sequential & 4 & 20 & 24000 & 0.47 & 0.36 & 0.46 & 3.19 & 0.40 & 1.7E+06\\
 &  &  &  & (0.24) & (0.14) & (0.07) & (5.31) & (0.001) & (3.0E+06)\\
KW & 4 & 20 & 24000 & 4.76 & 3.59 & 1.46 & 5.19 & 9.62 & 2.5E+03\\
 &  &  &  & (0.45) & (0.34) & (0.14) & (0.66) & (0.137) & (5.3E+03) 
  \end{tabular}
\caption{Estimation result 1 : Parameter estimation with another parameter set, and computational time and Bellman error. We replicate 200 times with generated observation. Values in the table are mean and standard deviation of the estimator from the replication.}
\label{estimation1}
\end{center}
\end{table}

\begin{table}[htbp]
\begin{center}
  \begin{tabular}{c||ccc|cccc|cc}
Method & $p$ & $T$ & $|\mathcal{S}|$ & $\theta_1$ & $\theta_2$ & $\theta_3$ & $\theta_4$ & Time (sec) & $\Delta^2$\\ \hline \hline
True param &  &  &  & 2.0 & 3.0 & 2.0 & 12.0&&\\ \hline
SLSTD & 4 & 10 & 3000 & 2.24 & 1.20 & 2.25 & 11.18 & 0.44 & 2.1E+02\\
 &  &  &  & (0.26) & (1.02) & (0.34) & (5.76) & (0.004) & (2.9E+02)\\
Sequential & 4 & 10 & 3000 & 1.96 & 2.99 & 2.10 & 11.73 & 0.09 & 2.1E+05\\
 &  &  &  & (0.03) & (0.02) & (0.01) & (0.19) & (0.002) & (2.9E+05)\\
KW & 4 & 10 & 3000 & 2.44 & 1.52 & 1.46 & -0.06 & 0.64 & 1.6E+03\\
 &  &  &  & (0.47) & (0.29) & (0.23) & (1.01) & (0.018) & (3.6E+03)\\ \hline
SLSTD & 4 & 15 & 10125 & 1.36 & 1.97 & 3.33 & 12.44 & 0.70 & 3.5E+03\\
 &  &  &  & (1.52) & (3.00) & (3.91) & (6.03) & (0.022) & (6.3E+03)\\
Sequential & 4 & 15 & 10125 & 0.19 & -0.45 & 1.32 & 10.18 & 0.16 & 2.7E+06\\
 &  &  &  & (1.10) & (3.10) & (3.40) & (15.33) & (0.001) & (4.7E+06)\\
KW & 4 & 15 & 10125 & 9.46 & 6.53 & -0.60 & 6.75 & 2.84 & 6.3E+03\\
 &  &  &  & (3.07) & (2.01) & (1.07) & (5.13) & (0.027) & (1.2E+04)\\ \hline
SLSTD & 4 & 20 & 24000 & 1.81 & 2.78	& 1.85 & 11.06 & 0.96 & 1.3E+04\\
 &  &  &  & (0.65) & (0.74)	& (0.50)	& (3.19) & (0.042)  & (2.8E+04)\\
Sequential & 4 & 20 & 24000 & 0.55 & 0.93 & 0.98 & 12.32 & 0.40 & 9.0E+06\\
 &  &  &  & (0.51) & (1.02) & (1.01) & (0.08) & (0.032) & (1.9E+07)\\
KW & 4 & 20 & 24000 & 10.39 & 3.71 & 4.57 & 6.17 & 9.61 & 1.1E+04\\
 &  &  &  & (0.28) & (0.52) & (1.04) & (6.10) & (0.069) & (2.5E+04)
  \end{tabular}
\caption{Estimation result 2 : Parameter estimation with another parameter set, and computational time and Bellman error. We replicate 200 times with generated observation. Values in the table are mean and standard deviation of the estimator from the replication.}
\label{estimation2}
\end{center}
\end{table}

\newpage

\begin{table}[htbp]
\begin{center}
  \begin{tabular}{c||ccc|cccc}
Method & $p$ & $T$ & $|\mathcal{S}|$ & Time (mean) & Time (s.d.)  \\\hline\hline
SLSTD & 3 & 20 & 1200 & 0.92 & 0.07  \\
Sequential & 3 & 20 & 1200 & 0.04 & 0.01 \\
KW & 3 & 20 & 1200 & 0.40 & 0.04 \\\hline
SLSTD & 3 & 30 & 2700 & 1.43 & 0.39 \\
Sequential & 3 & 30 & 2700 & 0.06 & 0.10 \\
KW & 3 & 30 & 2700 & 0.74 & 0.31  \\\hline
SLSTD & 3 & 40 & 4800 & 2.88 & 0.83 \\
Sequential & 3 & 40 & 4800 & 0.16 & 0.02 \\
KW & 3 & 40 & 4800 & 1.77 & 0.16 \\\hline
SLSTD & 4 & 20 & 24000 & 1.21 & 0.26 \\
Sequential & 4 & 20 & 24000 & 0.52 & 0.13 \\
KW & 4 & 20 & 24000 & 11.90 & 2.23 \\\hline
SLSTD & 4 & 30 & 81000 & 1.67 & 0.23 \\
Sequential & 4 & 30 & 81000 & 1.15 & 0.18 \\
KW & 4 & 30 & 81000 & 66.90 & 7.61 \\\hline
SLSTD & 4 & 40 & 192000 & 2.99 & 0.86 \\
Sequential & 4 & 40 & 192000 & 3.76 & 1.16  \\
KW & 4 & 40 & 192000 & 349.99 & 85.86 \\\hline
SLSTD & 5 & 20 & 480000 & 1.06 & 0.03 \\
Sequential & 5 & 20 & 480000 & 7.93 & 0.02 \\
KW & 5 & 20 & 480000 & 3081.82 & 178.30 \\\hline
SLSTD & 5 & 30 & 2430000 & 3.31 & 0.47  \\
Sequential & 5 & 30 & 2430000 & 69.55 & 6.82\\
KW & 5 & 30 & 2430000 & 82351.47 & 12344.05 \\\hline
SLSTD & 5 & 40 & 7680000 & 5.78 & 0.56 \\
Sequential & 5 & 40 & 7680000 & null & null \\
KW & 5 & 40 & 7680000 & null & null
\end{tabular}
\caption{Computation time (sec) : Computational time to solve some DMDs with difference size of state space. Values in the table are mean and standard deviation of the time length from the 200 replication.}
\label{time_all}
\end{center}
\end{table}

\newpage
\clearpage
\section{Proof of Lemma \ref{convlem}} \label{proof_conv}

In this proof, we keep $\theta$ fixed and omit the notation. As mentioned before, $V^*(s)$ is the solution of the Bellman equation, and $\phi^T(s)\hat{w}$ is the approximation value obtained by SLSTD method.

At the beginning, the approximation error of $\hat{V}(s)$ can be decomposed as
\begin{align*}
	||\phi^T(s)\hat{w} - V^*(s)||_{\mathcal{S}} \leq ||\phi^T(s)\hat{w} - \Pi_{\Phi}(s)V^*(s)||_{\mathcal{S}} + ||\Pi_{\Phi}(s)V^*(s) - V^*(s)||_{\mathcal{S}}.
\end{align*}

First, we consider the term $||\phi^T(s)\hat{w} - \Pi_{\Phi}(s)V^*(s)||_{\mathcal{S}}$. To evaluate this error, we have to show the existence of the optimal approximation weight $w^*$. Theorem 1 of \citet{RTW2002} shows that the Bellman equation of DMD models has a unique fixed point solution. Then, Lemma 6 in \citet{TR1997} guarantees the existence of an optimal $w^*$ that uniquely satisfies $\phi^T(s) w = \Pi_{\Phi}(s)\sum _a P(s|a;w)T[\phi^T(\cdot) w](s,a),\forall s$.

Next, we show that the sequence of $w$ generated by the stochastic approximation method converges to $w^*$. To show this, we verify the conditions of Theorem 2 in \citet{TR1997} and Theorem 17 in \citet{Benveniste2012}. Because the Bellman operation of DMD models is a contraction mapping, we can apply Lemma 9 of \citet{TR1997} and show that $(w - w^*)^T E_0[\sum _a \phi^T(s_t) (T[\phi'(\cdot)w_{t}](s,a) - \phi(s_t)w_{t})] < 0$. The existence of a stationary distribution is guaranteed by the combining of data. The compactness of $\mathcal{S}$ can satisfy the condition about the initial state. Then, we can apply the theory of \citet{TR1997}, and show that $\hat{w} \ra w^*$.

According to the discussion, we can evaluate the approximation error of the SLSTD. \citet{Tagorti2014} provide a theory and show that, with a large probability,
\begin{align*}
	||\phi^T(s)\hat{w} - \Pi_{\Phi}(s)V^*(s)||_{\mathcal{S}} = O\left(\frac{1}{\sqrt{N}}\log N\right).
\end{align*}
Here, $N$ is the maximum number of the iteration.

About the second term $||\Pi_{\Phi}(s)V^*(s) - V^*(s)||_{\mathcal{S}}$, this is an error of $l2$ projection of $V^*(s)$ onto the linear space spanned by the basis functions. When the domain of the function is continuous, this error is equivalent to the error of a least square series estimation, and \citet{Andrews1991} and \citet{Newey1997} provide the theoretical result for this estimation. By the assumption \ref{asm_conv}, most conditions of \citet{Newey1997} are satisfied. A rank condition of \citet{Newey1997} is a critical condition. To discuss about it, we denote $S_N = \{s :s \in \{s_i\}_{i=1}^N \}$ as a set of states observed in the set of transition. Since $q = |\mathcal{S}|$ increases at least order $N$, an i.i.d. data generating derives $|S_N| = O_P(N)$. Hence we can treat $N$ of \citet{Newey1997} and the number of transition as same. 
Then, we obtain
\begin{align*}
	||\Pi_{\Phi}(s)V^*(s) - V^*(s)||_{\mathcal{S}} &\leq ||\Pi_{\Phi}(s)\dot{V}^*(s) - \dot{V}^*(s)||_{\overline{\mathcal{S}}} \\
	&= O\left(\frac{1}{\sqrt{N}}\log(N) + \frac{k}{\sqrt{N}} +k^{\frac{1}{2}-\frac{m}{p}}\right).
\end{align*}

Thus, we obtain lemma \ref{convlem}.

\section{Proof of Theorem \ref{asymthm}} \label{proof_asym}

Let $\zeta _N := ||\phi^T(s)\hat{w}_{\theta} - V^*(s)||$. Then, Lemma \ref{convlem} gives $\zeta_N = O\left(\frac{1}{\sqrt{N}}\log N + \frac{k}{N^{1/2}} +k^{\frac{1}{2}-\frac{m}{p}}\right)$. Then, from the condition of Theorem \ref{asymthm}, it is easy to show that $\zeta _N = O(N^{-1/4})$ .

First, we show the consistency of $\hat{\theta}$. 
For the purpose, we will show the stochastic equicontinuity of $\mathcal{L}(\theta,V(\theta))$ in $\theta$, we evaluate
\begin{align*}
	&\mbox{Pr}(\sup_{\theta'} \sup_{\theta:||\theta-\theta_0||\leq \delta} |\mathcal{L}(\theta,\hat{V}(\theta) - \mathcal{L}(\theta',\hat{V}(\theta')))| > \epsilon),
\end{align*}
with some positive constant $\delta$ and $\epsilon$.
We decompose it as
\begin{align*}
	&|\mathcal{L}(\theta,\hat{V}(\theta) - \mathcal{L}(\theta',\hat{V}(\theta')))| \\
	&\leq |\mathcal{L}(\theta,\hat{V}(\theta) - \mathcal{L}(\theta',\hat{V}(\theta)))| + |\mathcal{L}(\theta',\hat{V}(\theta) - \mathcal{L}(\theta',\hat{V}(\theta')))| \\
	& \leq C_N^{(n)}||\theta - \theta'|| + |\mathcal{L}(\theta',\hat{V}(\theta) - \mathcal{L}(\theta',\hat{V}(\theta')))|,
\end{align*}
with some $C_n^{(1)} = O(1)$.
The last inequality comes from Assumption \ref{asm_asym}. Before considering the second term, we note the following inequality:
\begin{align*}
	&||\hat{V}(\theta) - \hat{V}(\theta')||_{\mathcal{S}}\\
	&\leq ||\hat{V}(\theta) - V^*(\theta)||_{\mathcal{S}} + ||V^*(\theta) - V^*(\theta')||_{\mathcal{S}} + ||V^*(\theta') - \hat{V}(\theta')||_{\mathcal{S}} \\
	&\leq ||V^*(\theta) - V^*(\theta')||_{\mathcal{S}} + o_P(1).
\end{align*}
The last inequality is from Lemma \ref{convlem}. 
By the definition of the $V(s;\theta)$, it is a weighted sum of the bounded utility function. Thus, we obtain
\begin{align*}
	||V^*(\theta) - V^*(\theta')||_{\mathcal{S}} \leq C_N' ||\theta - \theta'||,
\end{align*}
where $C_N' = O(1)$. This inequality is also obtained from Assumption 1. From the discussion, we can show
\begin{align*}
	||\hat{V}(\theta) - \hat{V}(\theta')||_{\mathcal{S}} \leq C_N' ||\theta - \theta'|| + o(1);
\end{align*}
Assumption \ref{asm_asym} enables us to show
\begin{align*}
	|\mathcal{L}(\theta',\hat{V}(\theta) - \mathcal{L}(\theta',\hat{V}(\theta')))| \leq C_N'' ||\theta - \theta'|| + o(1),
\end{align*}
where $C_N''=O(1)$.

Finally, we prove that for some $\delta >0$, there exists $\xi >0$ and
\begin{align*}
	&\lim_{\delta \ra 0} \limsup_{n \ra \infty} Pr(\sup_{\theta'} \sup_{\theta:||\theta-\theta_0||\leq \delta} |\mathcal{L}(\theta,\hat{V}(\theta) - \mathcal{L}(\theta',\hat{V}(\theta')))| > \epsilon) \\
	& \leq \lim_{\delta \ra 0} \limsup_{N \ra \infty} Pr(\sup_{\theta'} \sup_{\theta:||\theta-\theta_0||\leq \delta} C_N ||\theta - \theta'|| + C_N'' ||\theta - \theta'|| + o(1) > \epsilon) < \xi.
\end{align*}
Thus, we can obtain the stochastic equicontinuity of $\mathcal{L}(\theta,V(\theta))$. Using Assumption \ref{asm_asym} and Theorem in \citet{vdV2000}, we can show consistency $\|\hat{\theta} - \theta_0\| \ra 0$ as $N \ra \infty$.

To show the convergence rate and asymptotic normality, we apply Theorem of the semiparametric M-estimator in \citet{Kosorok2008}.
Since we obtain $\zeta _N = O(N^{-1/4})$, Assumption 2-\ref{asmp2_5} is sufficient to make $\mathcal{L}$ to be smooth.
Assumption 2 and the above result about the consistency satisfies the assumption of Theorem 21.1 in \citet{Kosorok2008}.
Then the result holds.




\begin{thebibliography}{26}
\providecommand{\natexlab}[1]{#1}

\bibitem[{Aguirregabiria and Mira(2002)}]{AM2002}
Aguirregabiria, V. and Mira, P. (2002) Swapping the nested fixed point
  algorithm: A class of estimators for discrete markov decision models,
  \textit{Econometrica}, \textbf{70}, 1519--1543.

\bibitem[{Aguirregabiria and Mira(2007)}]{AM2007}
Aguirregabiria, V. and Mira, P. (2007) Sequential estimation of dynamic
  discrete games, \textit{Econometrica}, \textbf{75}, 1--53.

\bibitem[{Andrews(1991)}]{Andrews1991}
Andrews, D.~W. (1991) Asymptotic normality of series estimators for
  nonparametric and semiparametric regression models, \textit{Econometrica},
  \textbf{59}, 307--345.

\bibitem[{Bajari \textit{et~al.}(2007)Bajari, Benkard and Levin}]{BBL2007}
Bajari, P., Benkard, C.~L. and Levin, J. (2007) Estimating dynamic models of
  imperfect competition, \textit{Econometrica}, \textbf{75}, 1331--1370.

\bibitem[{Benveniste \textit{et~al.}(2012)Benveniste, M{\'e}tivier and
  Priouret}]{Benveniste2012}
Benveniste, A., M{\'e}tivier, M. and Priouret, P. (2012) \textit{Adaptive
  Algorithms and Stochastic Approximations}, Springer Publishing Company,
  Incorporated.

\bibitem[{Bradtke and Barto(1996)}]{BB1996}
Bradtke, S.~J. and Barto, A.~G. (1996) Linear least-squares algorithms for
  temporal difference learning, \textit{Machine Learning}, \textbf{22}, 33--57.

\bibitem[{Dube \textit{et~al.}(2012)Dube, Fox and Su}]{DFS2012}
Dube, J.~P., Fox, J.~T. and Su, C.~L. (2012) Improving the numerical
  performance of static and dynamic aggregate discrete choice random
  coefficients demand estimation, \textit{Econometrica}, \textbf{80},
  2231--2267.

\bibitem[{Egesdal \textit{et~al.}(2015)}]{ELS2015}
Egesdal, M., Lai, Z., and Su, C. L. (2015). Estimating dynamic discrete‐choice games of incomplete information. \textit{Quantitative Economics}, \textbf{6}(3), 567-597.

\bibitem[{Gowrisankaran and Rysman(2012)}]{Gowrisankaran2012}
Gowrisankaran, G. and Rysman, M. (2012) Dynamics of consumer demand for new
  durable goods, \textit{Journal of Political Economy}, \textbf{120},
  1173--1219.

\bibitem[{Hendel and Nevo(2006)}]{HN2006}
Hendel, I. and Nevo, A. (2006) Measuring the implications of sales and consumer
  inventory behavior, \textit{Econometrica}, \textbf{74}, 1637--1673.

\bibitem[{Hotz and Miller(1993)}]{HM1993}
Hotz, V.~J. and Miller, R.~A. (1993) Conditional choice probabilities and the
  estimation of dynamic models, \textit{The Review of Economic Studies},
  \textbf{60}, 497--529.

\bibitem[{Ichimura and Lee(2010)}]{IL2010}
Ichimura, H. and Lee, S. (2010) Characterization of the asymptotic distribution
  of semiparametric m-estimators, \textit{Journal of Econometrics},
  \textbf{159}, 252--266.

\bibitem[{Imai and Keane(2004)}]{Imai2004}
Imai, S. and Keane, M.~P. (2004) Intertemporal labor supply and human capital
  accumulation, \textit{International Economic Review}, \textbf{45}, 601--641.

\bibitem[{Judd(1996)}]{judd1996}
Judd, K.~L. (1996) Approximation, perturbation, and projection methods in
  economic analysis, \textit{Handbook of computational economics}, \textbf{1},
  509--585.

\bibitem[{Keane and Wolpin(1994)}]{KW1994}
Keane, M.~P. and Wolpin, K.~I. (1994) The solution and estimation of discrete
  choice dynamic programming models by simulation and interpolation: Monte
  carlo evidence, \textit{The Review of Economics and Statistics}, \textbf{76},
  648--672.

\bibitem[{Keane and Wolpin(1997)}]{KW1997}
Keane, M.~P. and Wolpin, K.~I. (1997) The career decisions of young men,
  \textit{Journal of political Economy}, \textbf{105}, 473--522.

\bibitem[{Kosorok(2000)}]{Kosorok2008}
Kosorok, M. R. (2008). \textit{Introduction to empirical processes and semiparametric inference}. Springer Series in Statistics.


\bibitem[{Nedic and Bertsekas(2003)}]{Nedic2003}
Nedic, A. and Bertsekas, D.~P. (2003) Least squares policy evaluation
  algorithms with linear function approximation, \textit{Discrete Event Dynamic
  Systems}, \textbf{13}, 79--110.

\bibitem[{Newey(1997)}]{Newey1997}
Newey, W.~K. (1997) Convergence rates and asymptotic normality for series
  estimators, \textit{Journal of Econometrics}, \textbf{79}, 147--168.

\bibitem[{Pesendorfer and Schmidt-Dengler(2008)}]{PS2008}
Pesendorfer, M. and Schmidt-Dengler, P. (2008) Asymptotic least squares
  estimators for dynamic games, \textit{The Review of Economic Studies},
  \textbf{75}, 901--928.

\bibitem[{Pollard(1984)}]{Pollard1984}
Pollard, D. (1984) \textit{Convergence of Stochastic Processes},
  Springer-Verlag, New York.

\bibitem[{Rust(1987)}]{Rust1987}
Rust, J. (1987) Optimal replacement of gmc bus engines: An empirical model of
  harold zurcher, \textit{Econometrica}, \textbf{55}, 999--1033.

\bibitem[{Rust(1997)}]{Rust1997}
Rust, J. (1997) Using randomization to break the curse of dimensionality,
  \textit{Econometrica}, \textbf{65}, 487--516.

\bibitem[{Rust \textit{et~al.}(2002)Rust, Traub and Wozniakowski}]{RTW2002}
Rust, J., Traub, J.~F. and Wozniakowski, H. (2002) Is there a curse of
  dimensionality for contraction fixed points in the worst case?,
  \textit{Econometrica}, \textbf{70}, 285--329.

\bibitem[{Su and Judd(2012)}]{Su2012}
Su, C.~L. and Judd, K.~L. (2012) Constrained optimization approaches to
  estimation of structural models, \textit{Econometrica}, \textbf{80},
  2213--2230.

\bibitem[{Sutton(1988)}]{Sutton1988}
Sutton, R.~S. (1988) Learning to predict by the methods of temporal
  differences, \textit{Machine learning}, \textbf{3}, 9--44.

\bibitem[{Tagorti and Scherrer(2015)}]{Tagorti2014}
Tagorti, M. and Scherrer, B. (2015) Rate of convergence and error bounds for
  lstd({\(\lambda\)}),  \textit{In Journal of Machine Learning Research W \& CP, \textbf{37}(ICML 2015)}, 1521--1529.

\bibitem[{Tsitsiklis and Van~Roy(1997)}]{TR1997}
Tsitsiklis, J.~N. and Van~Roy, B. (1997) An analysis of temporal-difference
  learning with function approximation, \textit{Automatic Control, IEEE
  Transactions on}, \textbf{42}, 674--690.


\bibitem[{Tsybakov(2009)}]{Tsybakov2009}
Tsybakov, A. B. (2009). \textit{Introduction to nonparametric estimation}. Springer Series in Statistics.

\bibitem[{Van der Vaart(2000)}]{vdV2000}
Van der Vaart, A. W. (2000). \textit{Asymptotic statistics}. Cambridge university press.


\end{thebibliography}


\newpage
\end{document}